# Spin-dependent exciton quenching and intrinsic spin coherence in CdSe/CdS nanocrystals

Kipp J. van Schooten[1], Jing Huang[2], William J. Baker[1], Dmitri V. Talapin[2,3], Christoph Boehme[1*] and John M. Lupton[1,4*]

[1] Department of Physics and Astronomy, University of Utah, 115 South 1400 East, Salt Lake City, Utah 84112, USA

[2] Department of Chemistry, University of Chicago, Chicago, Illinois 60637, USA

[3] Center for Nanoscale Materials, Argonne National Laboratory, Argonne, Illinois 60439, USA

[4] Institut für Experimentelle und Angewandte Physik, Universität Regensburg, D-93040 Regensburg, Germany

**Large surface to volume ratios of semiconductor nanocrystals cause susceptibility to charge trapping, which can modify luminescence yields and induce single-particle blinking. Optical spectroscopies cannot differentiate between bulk and surface traps in contrast to spin-resonance techniques, which in principle avail chemical information on such trap sites. Magnetic resonance detection via spin-controlled photoluminescence enables the direct observation of interactions between emissive excitons and trapped charges. This approach allows the discrimination of two functionally different trap states in CdSe/CdS nanocrystals underlying the fluorescence quenching and thus blinking mechanisms: a spin-dependent Auger process in charged particles; and a charge-separated state pair process, which leaves the particle neutral. The paramagnetic trap centers offer control of energy transfer from the wide-gap CdS to the narrow-gap CdSe, i.e. light harvesting within the heterostructure. Coherent spin motion within the trap states of the CdS arms of**

[*] Corresponding authors. Email: john.lupton@physik.uni-regensburg.de ; boehme@physics.utah.edu

**nanocrystal tetrapods is reflected by spatially remote luminescence from CdSe cores with surprisingly long coherence times of >300 ns at 3.5 K.**

Substantial control over the chemistry of semiconductor nanocrystals has been demonstrated in recent years while pursuing novel optoelectronic device schemes[1,2,3]. Shortcomings in the performance of these materials are routinely attributed to ill-defined "trap" states competing with the quantum-confined primary exciton[4]. While frequently implicated in explaining device inefficiencies[2], photoluminescence (PL) blinking[5-9] and delayed PL dynamics[4,10], little is known about the underlying chemical nature of these deleterious states. Despite the wealth of structural and electronic information accessible in optical spectroscopy, the spin degree of freedom has received only marginal consideration as a complementary probe of semiconductor nanocrystals. Approaches pursued previously include isolation of paramagnetic centers in doped dilute magnetic semiconductor nanoparticles[11-12]; resolving the exciton fine structure by fluorescence spectral line narrowing[13], time-resolved Faraday rotation[12,14] or photon-echo techniques[15]; and continuous-wave optically-detected magnetic resonance (ODMR), where the fluorescence is modulated under spin-resonant excitation in a magnetic field[16-19]. The latter requires stable paramagnetic centers, where the carrier's spin and energy are maintained on long timescales compared to the oscillation period of the resonantly driven spin manifold, i.e. for tens of nanoseconds under excitation in the 10 GHz (~0.3 T) range. The persistence of spin states in bulk materials comprising heavy atoms such as cadmium is largely determined by mixing due to spin-orbit coupling.

As dimensions shrink to quantum-confined regimes, spin-orbit driven spin-mixing mechanisms can be weakened by the discretization and separation of states, giving way to the more subtle Fermi-contact hyperfine mode of spin mixing[20]. Although spin stability can be reinforced through quantum confinement, direct band-edge excitons in nanocrystals typically decay within a few nanoseconds, making them unsuitable for spin-resonant manipulation. In fact, spin mixing amongst the fine-structure levels[14,21] of excitonic states has been shown to occur within as little as a few hundred femtoseconds by means of photon-echo spectroscopy[15]. However, electronic charge-separated or "shelved" states also exist, where the excitonic constituents – either electron or hole, or both – are stored within a trap. The carriers in this case are not necessarily lost to non-radiative relaxation, but can feed back into the exciton state at a later time. A direct visualization of this phenomenon is given by the ability to store excitons in nanoparticles under an electric field[10,22], in analogy to excitonic memory effects in coupled quantum wells[23]. These charge-separated states can repopulate the exciton, since luminescence returns in a burst following field removal[10,22]. While qualitative information on these shelving states (which are distinct from chemical deep traps with their characteristic red-shifted emission with respect to the exciton) continues to feed the proliferation of microscopic models of quantum dot blinking[4-9,24,25], a more quantitative metrology is required to determine the nature and location of trapped charges. Such an approach is given by the highly-sensitive method of *pulsed* ODMR spectroscopy, which, in principle, is capable of chemically fingerprinting even single carrier spins.

We focus on the spin dynamics in CdSe/CdS nanocrystal tetrapods since absorption and emission can be well separated spatially and energetically: at 3.1 eV (400 nm), the absorption cross-section of the CdS arms is more than 300 times greater than that of the CdSe core[1]. Emission from CdSe dominates due to the lower bandgap, making the structures excellent light-harvesting systems[1,26]. Fig. 1 illustrates the underlying scheme. Photons are absorbed in the arm, leading to bright CdS excitons. The conduction bands of CdS and CdSe are approximately aligned, whereas a step of ~0.7 eV exists between the valence bands. We note that significant heterogeneity in the precise energetics of the heterostructure arises between single particles[26,27]. The direct transfer of CdS excitons to CdSe is not suspected to be spin dependent since CdSe[21] and CdS[28] ground-state exciton fine-structure should be the same for the size of nanocrystals used here. Further, energy transfer proceeds so rapidly as to inhibit spin manipulation. However, trap states for CdS excitons also exist, the influence of which is clearly seen in delayed PL (Fig. S1) where shelved excitons feed back into band-edge states at times much longer than the exciton lifetime. We therefore manipulate the spin state of charge pairs shelved within the CdS, provided these maintain their spin identity while trapped. We do not directly manipulate those spins corresponding to the band-edge exciton fine-structure. Spin resonance can then induce a conversion of mutual spin orientation for trapped carrier (electron-hole) pairs, converting them from "bright" to "dark" permutation symmetry. Once detrapping occurs (as in the delayed PL in Fig. S1), these weakly (exchange and magnetic dipole-) coupled spin pairs again become strongly-coupled band-edge exciton states where the mutual spin identity of the trapped carriers largely predetermines which excitonic fine-structure level becomes populated. Since the trap energies we concern ourselves with are, or are nearly, iso-energetic with the band-gap, spin-scattering while moving in and out of trap states is weak. This process of cycling carriers from band-edge excitons to traps, changing trap-state spin configuration, and then moving the carriers back to excitonic states is what generally allows spin-dependent PL in our structures: dark shelved carriers determine the population ratio for dark band-edge excitons, which remain dark upon transfer to the CdSe core of the nanocrystal. It is important to note here that since PL is the observable in this scheme, at this time a direct discrimination cannot be made between scenarios involving PL quenching due to an increase in trapping lifetime or quenching due to a direct transfer into a dark exciton[14,21] state. In either case, the bright exciton population is diminished.

Fig. 1b) summarizes the experimental approach (full details are provided in the *Supporting Information*). A sample of tetrapods is illuminated by a continuous-wave laser, and a homogeneous magnetic field splits the Zeeman sublevels. Transitions between these levels are induced coherently during the application of microwaves and, for optically-active charge carriers, this process is witnessed as a transient perturbation in PL intensity with respect to the steady-state. A typical luminescence transient is illustrated in the figure: the microwave pulse should lead to luminescence *quenching* since optical excitation initially populates bright exciton states[21], but coherent spin mixing of intermediately shelved carriers leads to an overall increase of dark state exciton populations. After removal of the microwave field, the PL intensity returns slowly as shelved "dark" states undergo spin-lattice relaxation to form "bright" configurations which feed back into bright band-edge excitons. This longer timescale process can result in an eventual *enhancement* over the steady-state background as long as the intersystem crossing rate is low relative to the rate of initial PL quenching[29]. The resonances of the composite CdSe/CdS material are surveyed in Fig. 2. In order to fully identify the material and spectral origin of observed resonant species, we compare separately CdSe quantum dots, CdS nanorods, and the full composite CdSe/CdS tetrapod heterostructures. The CdSe core emission spectrum is shown in Fig. 2a). The corresponding ODMR spectrum (panel b), where the differential PL is plotted as a function of magnetic field and time after the microwave pulse, shows only weak PL enhancement and no quenching, exhibiting broad inhomogeneity. We tuned the magnetic field over 1 T and found continuous PL enhancement over a range of 500 mT. The broad resonance is attributed to deep (below 2.1 eV) red-emitting highly spin-orbit coupled chemical defects of CdSe, and not to the band-edge exciton[16,17].

CdS nanorods are also known to emit at two energies; at ~2.667 eV (465 nm) due to the quantum-confined band-edge exciton, and in a broad spectrum around 2.066 eV (600 nm) due to a deep-level chemical defect associated with a surface sulfur vacancy. The features are seen in the emission spectrum in panel c). The ODMR transient mapping of the defect emission (selected by an emission filter) is illustrated in panel d). A resonance is identified at 352 mT, corresponding to *enhancement* of defect PL, which decays over ~50 μs. In contrast, detection in the narrow exciton band (emission filter region marked blue in panel c) reveals distinct behavior (panel e): two resonances dominate, at 345 mT and 374 mT, corresponding to PL *quenching* under resonance. After ~30 μs, PL enhancement occurs. As discussed above, this transient interplay of PL quenching and enhancement is as expected for band-edge trap states experiencing slow intersystem crossing and intermixing with exciton states. In the following, we focus only on resonances associated with the exciton emission channel rather than luminescence of the defect, since the former likely relate to traps responsible for single-particle blinking[7-9]. As outlined below (and further in the *Supporting Information*), the two band-edge resonances arise due to a pair of weakly-coupled spin-½ species, i.e. electron and hole. In contrast to the bare cores (panels a-b), the same CdSe emission spectrum measured from the tetrapods (panel f) shows ODMR characteristics that are dominated by the *CdS* band-edge trap states (panel g). Here, spin-dependent transitions of the CdS are imparted on the core emission, enabling remote readout of CdS arm spin states. Such ODMR signals were only observed at low temperatures, their amplitude increasing steadily from 50 K down to 3.5 K.

To clarify the origin of spin-dependent transitions in the CdS exciton emission we inspect the resonance dynamics. The nanorod ODMR spectrum in Fig. 3a), recorded 3.2 μs after a microwave pulse of 800 ns duration (i.e. a vertical slice of Fig. 2e), is accurately described by the sum of three Gaussian resonances. One peak is located at a characteristic Landé $g$-factor of $g = 2.0060(2)$ (blue arrow), suggesting that this resonance is related to a semi-free charge [$g_{free-electron} \sim 2.002319$] with negligible spin-orbit coupling. The second distinct peak [black arrow, $g = 1.8486(2)$] is substantially shifted from the free-electron value, indicating that the carrier is localized in a trap with significant spin-orbit coupling. The third Gaussian is environmentally broadened (i.e. by hyperfine fields and a variation in effective spin-orbit coupling) and centered at $g = 1.9594(2)$ (grey arrow). Panel b) plots the absolute differential PL against time after resonant microwave excitation for the black and blue peaks, revealing that the perturbed spin-state populations follow identical time dynamics during free spin evolution. The decay of the pronounced initial quenching signal approximately follows a single exponential, indicating a dominant single spin-dependent transition rate. This transient is succeeded by a long-term PL enhancement, again dropping exponentially between 300-800 μs after microwave excitation. This form of transient, involving two primary exponential rates, is a clear signature of an electron-hole pair process[30]. Nearly identical resonance line shapes and dynamics are extracted for the tetrapods (panels c, d), confirming that spin information existing in the CdS nanorods can indeed be accessed via luminescence from the attached CdSe core. As seen in the comparison between the two sets of nanoparticles, line shapes and resonance center positions are expected to be subject to minor variations since both size and geometry of the particles affect quantum confinement and therefore the relative $g$-factors[14,21,28]. On average, the differential PL is ten times weaker for the tetrapods than for the nanorods, since light-harvesting of the CdS excitons inhibits trapping on metastable sites as required for this spin-resonant manipulation.

The resonances around $g \sim 2.00$ and $g \sim 1.84$ not only follow the same decay to equilibrium after spin-mixing, but the spectral integrals also match (Fig. S2). This agreement is expected for a correlated spin-½ pair process (Fig. S3); manipulation of either electron *or* hole spin has equal probability of modulating PL since the two charges couple by the same spin-dependent mechanism. An intriguing conclusion can be drawn from these observations: the band diagram of the tetrapods in Fig. 1 suggests that the hole should immediately localize in the CdSe core, although this is obviously not the case[31]. If this were the case, we would not observe identical resonance spectra and dynamics in nanorods and tetrapods. Instead, for the same spin-resonant manipulation of *both electron and hole* to occur in the tetrapods, both must be located within the CdS on the *same* nanoparticle and at the *same* time. The ODMR data therefore imply that trapping of both charge carriers can occur simultaneously at the band-edge, a result that may be related to the recent spectroscopic identification of interfacial barriers at the CdSe/CdS interface[26]. Without significant modifications to the measurement technique or access to exact chemical information of at least one site, we are unable to assign a particular charge to these trap states since spin resonance techniques are inherently insensitive to the sign of a charge.

Given the lack of spin-orbit coupling (a shift from the free-electron g-value) and only limited environmental broadening, we propose that the $g \sim 2.00$ peak originates from a "semi-free" charge localized to the surface of the nanocrystal. The $g \sim 1.84$ resonance is only slightly broader than the $g \sim 2.00$ line, indicating that it is also associated with a localized surface site rather than the bulk, but is shifted due to spin-orbit coupling. We note that a resonance near $g \sim 2.00$ has previously been reported[32] for photogenerated holes in CdS[33], but this is also the expected *g*-value for charges localized to organic ligands[34] or matrix material[5] experiencing negligible spin-orbit coupling. This type of interaction with surface ligands is a distinct possibility as is evidenced by the lack of a phonon bottleneck in colloidal quantum dots, a phenomenon which has been shown to be mediated by carrier wavefunction overlap with organic ligands[35,36]. At present, the information needed to precisely discriminate between these two chemical situations is not complete (see *Supporting Information* for discussion). The $g \sim 1.84$ feature is distinct from that found in ODMR of bulk CdS[37] ($g = 1.789$), although *g*-factors can shift significantly due to quantum size effects and geometry[14,21,28].

The third feature, the broad $g = 1.95$ peak marked grey in Fig. 3, only shows quenching and no enhancement, and decays faster than the narrow resonances, demonstrating that it arises from a distinct spin-dependent process (Fig. S2). This feature vanishes in the tetrapods for excitation below the CdS band gap (Fig. S5). The broadening is likely induced by local strain or hyperfine fields, or by a superposition of multiple unresolved resonances. We propose that the resonance originates from a carrier trapped *within* the nanocrystal where a wide range of *g*-factors exists. This ODMR signal then likely arises due to spin-dependent Auger recombination[30] between the localized carrier and the quantum-confined band-edge exciton within the particle, a process known to quench optical recombination[5,7-9].

Rapid spin dephasing would normally be anticipated for a bulk-like crystal, given the significant spin-orbit coupling of the $g \sim 1.84$ resonance[15]. However, recording differential PL at each distinct resonance as a function of microwave pulse duration reveals Rabi flopping as displayed in Fig. 4), a direct manifestation of spin-phase coherence. In this example, spins precess so that the shelved carrier pairs propagate *reversibly* between bright and dark mutual spin configurations. Such Rabi oscillations were recently reported for Mn-doped CdSe nanocrystals by conventional absorptive magnetic resonance[11], but are unprecedented for direct detection via intrinsic optical transitions of the semiconductor. The frequency components contained within this coherent oscillation provide additional information on the nature of these states; specifically on carrier spin-multiplicity and the existence of exchange and/or dipolar coupling. From this analysis (a detailed treatment is given in the *Supporting Information*), it is found that both the $g \sim 2.00$ and $g \sim 1.84$ resonances describe carriers which carry spin-½. The mutual exchange and dipolar coupling experienced within the trapped pair is negligible.

Although the decay of the Rabi oscillation can provide a lower bound on the coherence lifetime for each of these carriers, more sophisticated resonant-pulse sequences can be used to unambiguously measure this value. We quantify the CdS spin-phase lifetime, $T_2$, by measuring Hahn spin echoes, the amplitude of differential PL change following rephasing of spins by a second microwave pulse (Fig. S4). Fig. 4a),b) (insets) exhibit exponential decay of the echo amplitude as a function of interpulse delay τ, yielding $T_2 =328\pm22$ ns for $g \sim 2.00$ and $186\pm12$ ns for $g \sim 1.84$. The coherence time of the $g = 1.95$ resonance is too short to be measured using our technique ($T_2 <$ several ns). Additional structure is seen on the echo decay of the $g \sim 2.00$ carrier due to hyperfine-field-induced electron spin-echo envelope modulation[11] (see *Supporting Information*). The pair partner of this quasi-free carrier, localized to a surface trap, experiences stronger spin-orbit coupling, lowering the *g*-factor and accelerating dephasing. Nevertheless, these $T_2$ values are unprecedented for non-magnetic semiconductor nanocrystals[11,38].

It is notable that, as a consequence of these extraordinary coherence times, identical Rabi oscillations result under detection of CdS (nanorods, Fig. 4b) and CdSe emission (tetrapods, Fig. 4c). The experiments offer a qualitative assessment of the degree of trap localization. This must be significant since delocalized carriers would be expected to lose coherence by coupling to a new environment, such as the core of the tetrapods. The persistence of spin coherence over different system environments offers the possibility of remote readout of spin-phase information, and demonstrates the fundamental ability to coherently control light-harvesting[39] even in inorganic structures.

Pulsed ODMR directly reveals three radical species in CdS which control PL and are likely responsible for the two types of blinking observed in CdSe/CdS particles as distinguished by luminescence lifetime[8]: either both carriers are localized to the nanocrystal surface, leaving the particle neutral and thus preventing Auger recombination and a change in exciton lifetime ($g \sim 2.00$ and $g \sim 1.84$); or one carrier is trapped *within* the particle ($g = 1.95$), charging it so that Auger-type blinking with the associated fluorescence lifetime changes arises. This localization of carriers occurs in CdS, not CdSe. Surprisingly, shelved excitons do not thermalize from CdS to CdSe, but remain in the CdS "shell" of the heterostructure nanoparticle[31]. The extraordinarily long spin quantum-phase coherence times of order $1\mu s$ highlight the potential utility of even strongly spin-orbit coupled nanoparticles for quantum information processing or quantum-enhanced sensing such as magnetometry. In contrast to conventional inorganic quantum systems, such as electrostatically-defined quantum dots, nanocrystals offer the possibility of creating spatially-scalable quantum structures through bottom-up synthetic means[3] as demonstrated here by the spatially remote light-harvesting read-out of spin-phase information.

**Acknowledgements**

Acknowledgment is made to the Department of Energy Grant (#DESC0000909) for funding of this research. We thank the National Science Foundation for support through MRSEC Projects #1121252 and #0213745. J.M.L. and D.V.T. are indebted to the David and Lucile Packard Foundation for providing fellowships. C.B. and D.V.T. acknowledge support by National Science Foundation CAREER grants (#0953225 and #0847535, respectively).

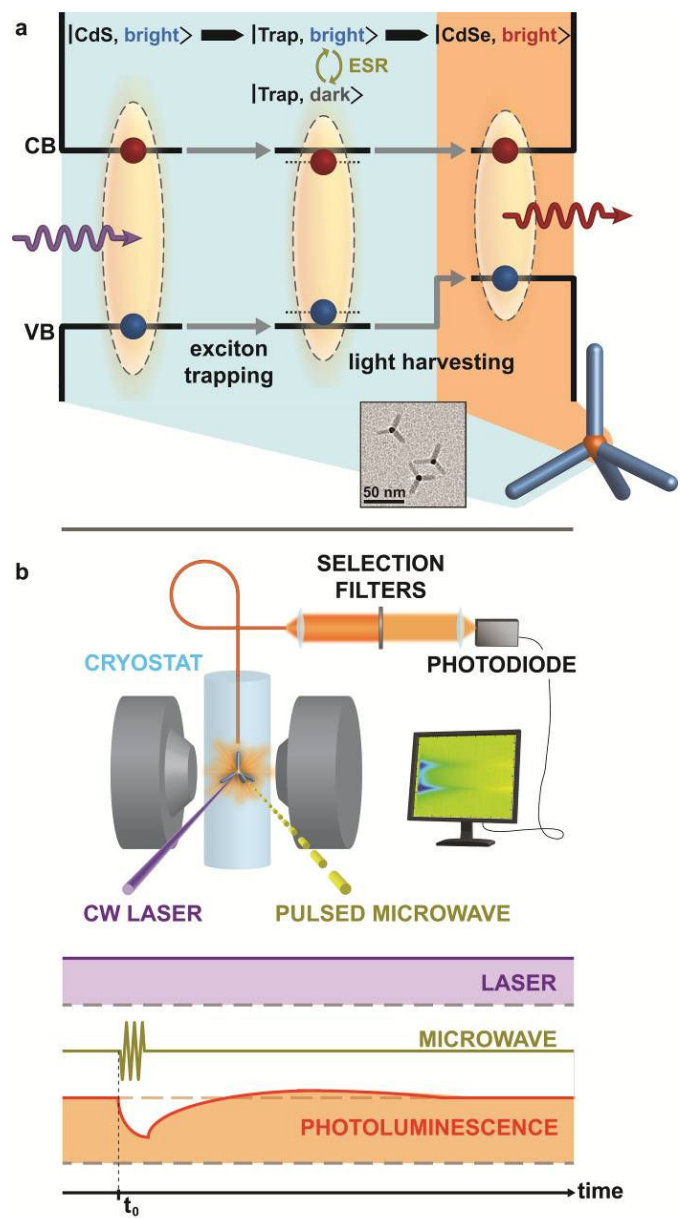

**Figure 1. Pulsed optically-detected magnetic resonance of light-harvesting tetrapod CdSe/CdS semiconductor nanocrystals.** a) Excitons are generated in the CdS arms by light absorption. A small fraction of these excitons becomes trapped as charge-separated states, which can reemit an exciton to the CdS band edge. The lifetime of the trapped state is sufficient to enable spin manipulation via electron spin resonance (ESR), switching the trapped carrier pair between "bright" and "dark" mutual spin

configurations. Relaxation of the exciton to the CdSe core gives rise to strongly red-shifted emission. The transmission electron micrograph inset illustrates the high quality of the structures used. b) Experimental setup and a representative differential PL transient as a consequence of resonant spin transition of an optically active carrier.

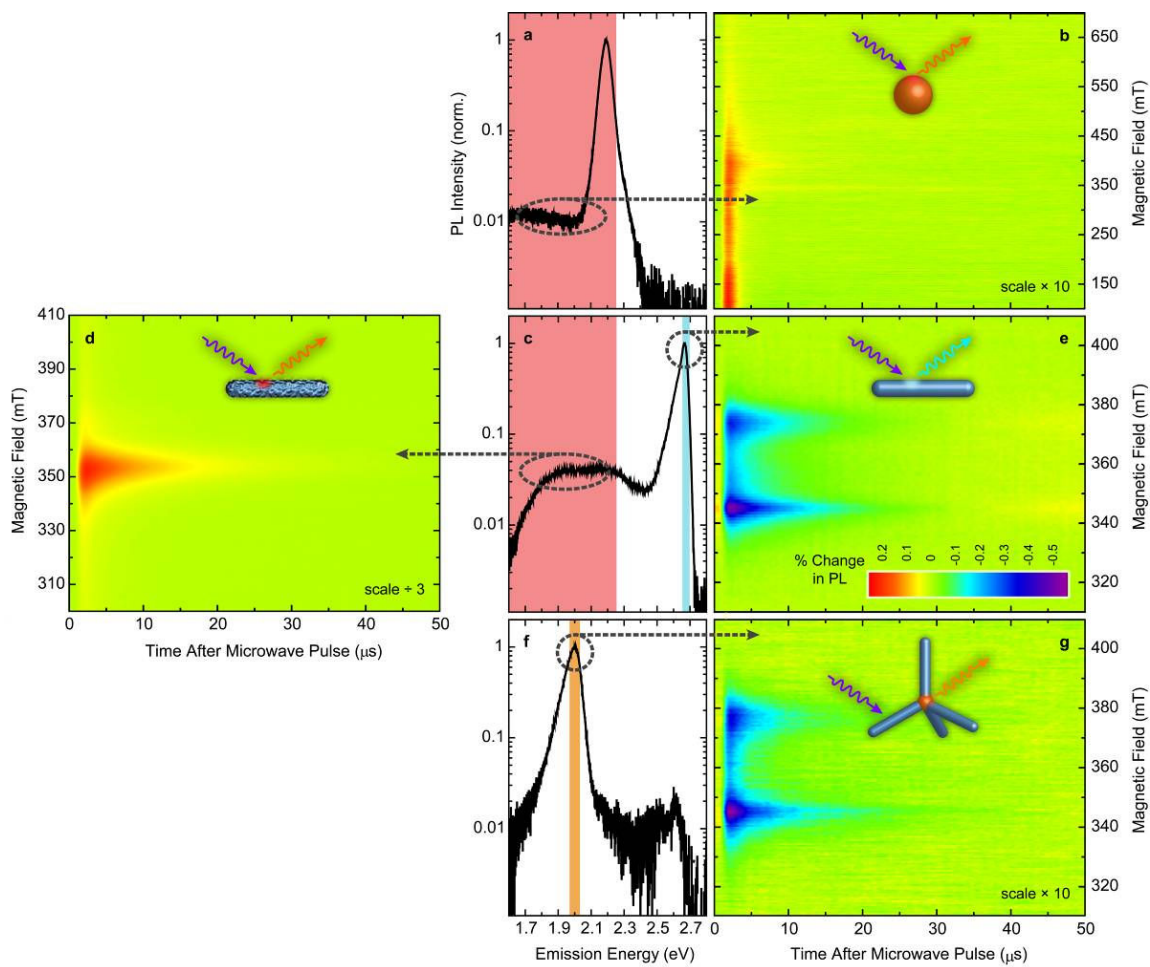

**Figure 2. Optically-selected spin-dependent transitions in semiconductor nanocrystals at 3.5 K under X-band (9.8 GHz) excitation.** a) Emission spectrum of bare CdSe nanocrystal quantum dots (tetrapod cores) and associated near-featureless transient ODMR spectrum (b) taken as a function of emission intensity (filter region marked in red) in dependence of magnetic field following a microwave pulse. c) Emission spectrum of CdS nanorods. d) Differential PL (enhancement) of CdS nanorod deep-trap level defect emission (marked red in panel c). e) Differential PL (quenching) of the CdS band-edge exciton emission (band labeled blue in panel c). f) PL spectrum of CdSe/CdS nanocrystal tetrapods with associated transient ODMR spectrum detected in the CdSe emission (g), revealing the CdS spin species. The colored bars in panels a), c),

f) indicate the spectral region of the transmission filters used. The laser excitation energy is chosen to be just above the CdS nanorod bandgap (~2.7 eV).

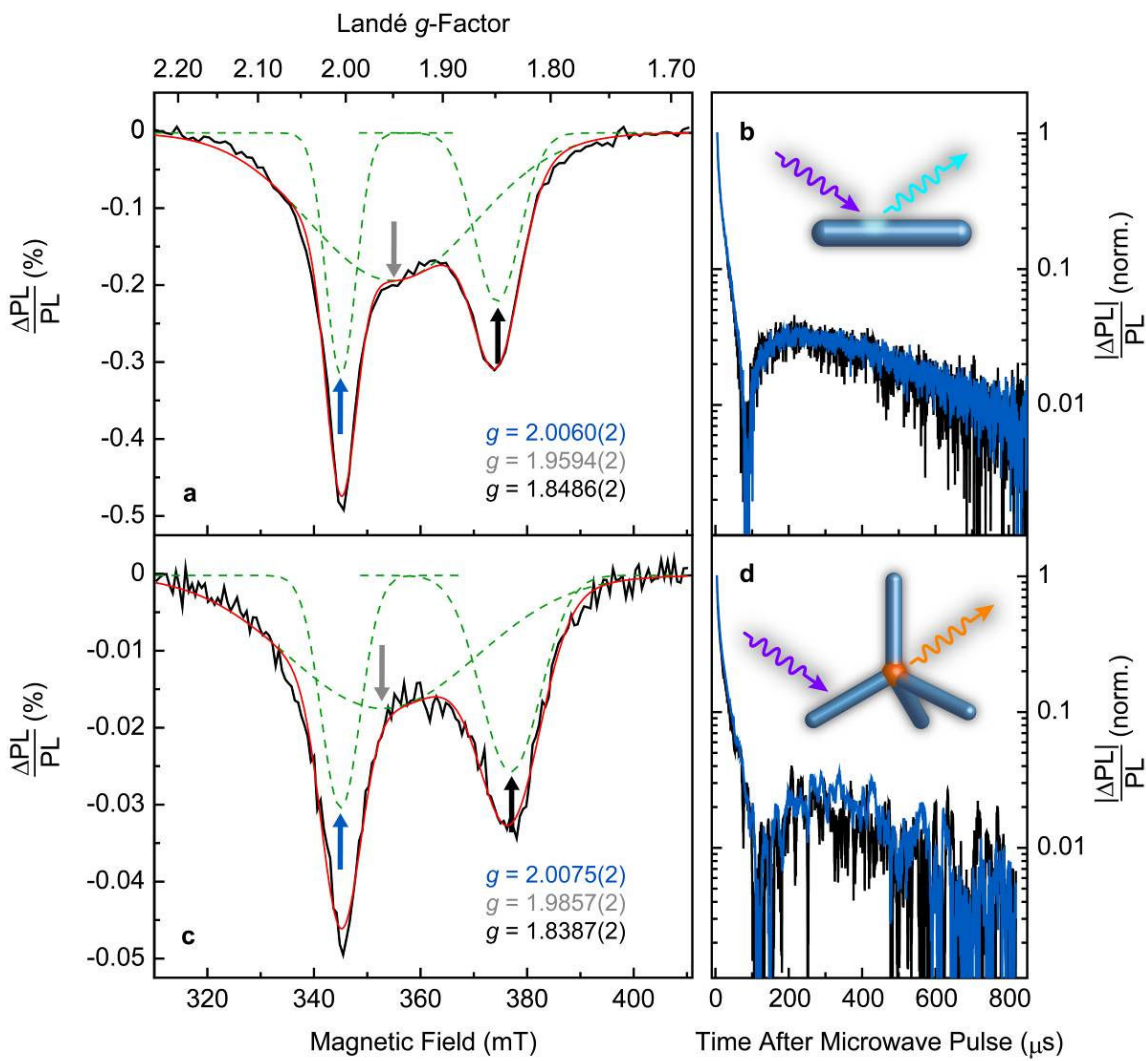

**Figure 3. Dynamics of distinct spin-resonant species in ODMR.** Spectra at 3.2 μs delay following an 800 ns microwave pulse detected in the band-edge emission of CdS nanorods (a) and in the CdSe core emission of tetrapods (c). The spectra are accurately described by a superposition of three Gaussians. The temporal dynamics of the two dominant resonances (marked blue and black) are identical in b) and d), implying that the two spin-½ species are correlated. We assign these peaks to spin dynamics in a charge-separated state, with each charge carrier located on the surface of the nanocrystal. The charge at $g \sim 2.00$ represents a "semi-free" carrier while the pair partner ($g \sim 1.84$) is situated on a site with greater spin-orbit interactions. The third broad Gaussian resonance follows different temporal dynamics and originates from an unrelated trapped species, located *within* the CdS where a large distribution in resonance frequencies exists. We tentatively assign the single broad resonance to a spin-dependent Auger-type process and the pair mechanism to the situation where both carriers are expelled from the bulk of the particle, generating surface charge which modulates fluorescence but leaves the particle neutral. Light harvesting in the tetrapods reduces the number of shelved excitons since carriers are rapidly removed from the CdS, leading to a tenfold reduction in signal strength.

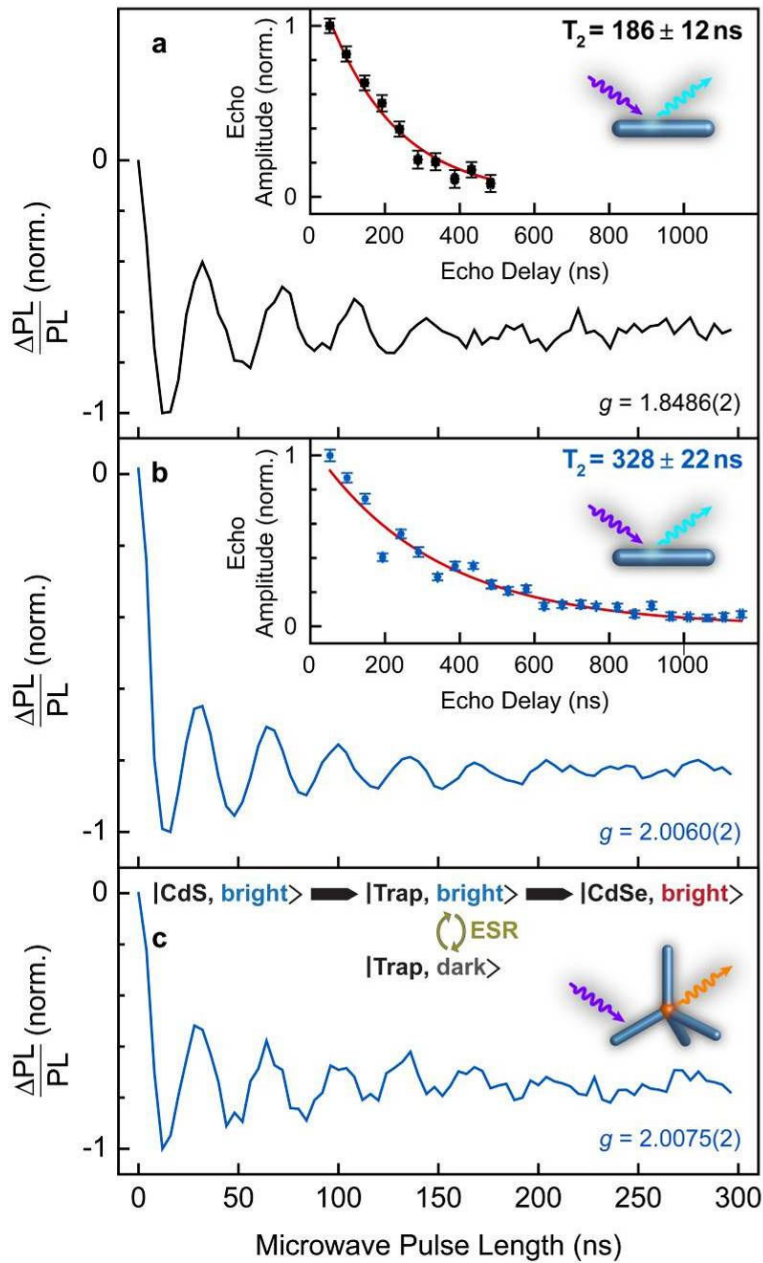

**Figure 4. Spin dephasing and coherent control of light harvesting in nanocrystal tetrapods.** (a), (b) Rabi oscillations in the differential PL of the CdS nanorods as a function of microwave pulse length for the resonances around $g \sim 1.84$ and $g \sim 2.00$. The insets show the corresponding decay of spin coherence measured by performing Hahn spin echoes using a sequence of microwave pulses. (c) In the tetrapods, coherent

spin information in the CdS is extracted remotely in the PL of the CdSe core, indicating the high degree of carrier localization since coherence information remains unperturbed upon change of environment (i.e. addition of the core to form the tetrapod heterostructure). This result also demonstrates the ability to coherently control the light-harvesting process.

# Supporting Information for:

# Spin control of emission-detected light harvesting in CdSe/CdS nanocrystals due to long intrinsic quantum coherence

K. J. van Schooten[1], J. Huang[2], W. J. Baker[1], D. V. Talapin[2, 3], C. Boehme[1] * and J. M. Lupton[1, 4] *

[1] Department of Physics and Astronomy, University of Utah, 115 South 1400 East, Salt Lake City, Utah 84112, USA

[2] Department of Chemistry, University of Chicago, Chicago, Illinois 60637, USA

[3] Center for Nanoscale Materials, Argonne National Laboratory, Argonne, Illinois 60439, USA

[4] Institut für Experimentelle und Angewandte Physik, Universität Regensburg, D-93040 Regensburg, Germany

## Experimental Methods

The tetrapod nanocrystals consist of wurtzite CdS arms approximately 20 nm in length and 6 nm in diameter, grown onto four faces of zincblende CdSe cores of 4 nm diameter. Synthesis details are given in Ref. 1. The same batch of CdSe cores which was used to seed tetrapod growth was also investigated alone for comparison, as shown in Figure 2b) of the main text. Each series of colloidal nanoparticles used in these measurements was first diluted into a toluene Zeonex (Zeon Chemicals L.P.) solution and then drop cast into a small Teflon bucket (2mm × 3mm). Upon solvent evaporation, a solid matrix was formed, which is both optically and paramagnetically inert, but contains the distributed nanoparticles. The sample was then suspended in a He flow cryostat containing a dielectric microwave resonator, generally kept at 3.5K for all measurements, except for Rabi nutation experiments which was performed at 15K. Optical access to the sample was made by extending a home-built fiber bundle through a cryostat port and into the resonator, resting at the mouth of the Teflon sample bucket. A c.w. Ar$^+$ laser, tuned to 457.9 nm (2.708 eV) and combined with a suitable filter to remove spontaneous emission (Semrock Maxline), was passed into a single fiber and used to excite the nanocrystal ensemble with 20mW of power (intensity approximately 85µW·cm$^{-2}$). The remainder of the fibers were used to collect PL, from which scattered laser light was filtered out with a 458 nm ultrasteep long-pass filter (Semrock RazorEdge). Specific emission bands for each nanoparticle ensemble were spectrally selected by choosing an appropriate filter set: the CdS nanorod and CdSe core deep-level defect emission were isolated with a 550 nm (2.254 eV) long-pass filter (ThorLabs); the CdS nanorod band-edge emission was cut with a 460±2 nm (2.695±0.012 eV) narrow-band filter (ThorLabs); the tetrapod core emission was picked with a 620±2 nm (2.000±0.007 eV) narrow-band filter (ThorLabs).

---

* Corresponding authors. Email: john.lupton@physik.uni-regensburg.de ; boehme@physics.utah.edu

The selected PL was focused onto a low-noise photodiode (Femto LCA-S-400-Si), whose signal was amplified with a Stanford Research Systems low-noise preamplifier (SR560). AC coupling of the input signal was used in order to apply gain to only the modulated contribution of the PL intensity. A 300Hz high-pass frequency filter was also applied in order to help isolate the transient response of the ODMR signal from spurious electrical and optical modulations.

With sufficient gain applied, the resulting signal was passed into the fast digitizer of a Bruker SpecJet contained within an Elexsys E580 system, which correlates the timing of the microwave pulse sequence with the transient response. Programmable control over pulse routine timing, leveling of the external magnetic field and signal acquisition was utilized to carry out the large number of measurements required for each data set. For example, the transient mappings displayed in Fig. 2c),d),f), and h) required an X-band (9.8GHz) microwave pulse of 800ns duration to be applied every 800µs a total of 16384 times. The transient responses of the individual measurements were added together before incrementing the external magnetic field $B_0$. For the high-resolution time transients shown in Fig. 3b),d), the microwave shot repetition rate was set to be greater than 2ms, much longer than the full relaxation time to steady-state populations of carrier states under constant excitation of the material system. Rabi oscillations were obtained by monitoring the amplitude of the transient PL response as a function of microwave pulse length. The transit times for driving the system from optically dark to optically bright states served as useful timing information needed for constructing the π and π/2 pulses of the Hahn echo sequence. A full description of this conventional pulse sequence, as used in ODMR, is given below.

**Band-edge trap states in CdS nanocrystals**

The existence of trap states lying very close to the band gap of our primary material system of interest, the CdS nanorods, can easily be confirmed by considering the luminescence decay characteristics following an optical excitation pulse. A sample similar to that used for the ODMR experiments is fabricated, consisting of nanorods suspended in a polystyrene block several microns thick. This sample is mounted, under vacuum, to the cold-finger of a closed-cycle Helium cryostat, which cools to 21 K. A diode laser operating at 355 nm (3.493eV) with nanosecond pulse length and variable repetition rate is used as an excitation source. PL spectra are monitored with a gated, intensified CCD (ICCD) camera mounted to a spectrometer, allowing us to record the decay of emission intensity as a function of gating time following optical excitation. The prompt PL is shown in Figure S1. The dashed green line in panel a) indicates the spectral position which is monitored as a function of time. As is seen in panel b), the PL intensity drops off approximately following a power law over five orders of magnitude in time. The excitonic emission spectrum does not shift significantly over this time.

The accepted physical mechanism responsible for delaying emission in these nanoparticles for such long times is the temporary isolation of the optically excited charge carriers into their respective "trap" states[1], dramatically decreasing the amount of wavefunction overlap of the electron-hole pair, and therefore the likelihood of recombination. As the detrapping rate back into the band-edge excitonic states depends exponentially on the trap energy, which in turn is distributed exponentially, a distribution of detrapping rates is observed across the nanoparticle ensemble, leading to the power law-like emission decay[2].

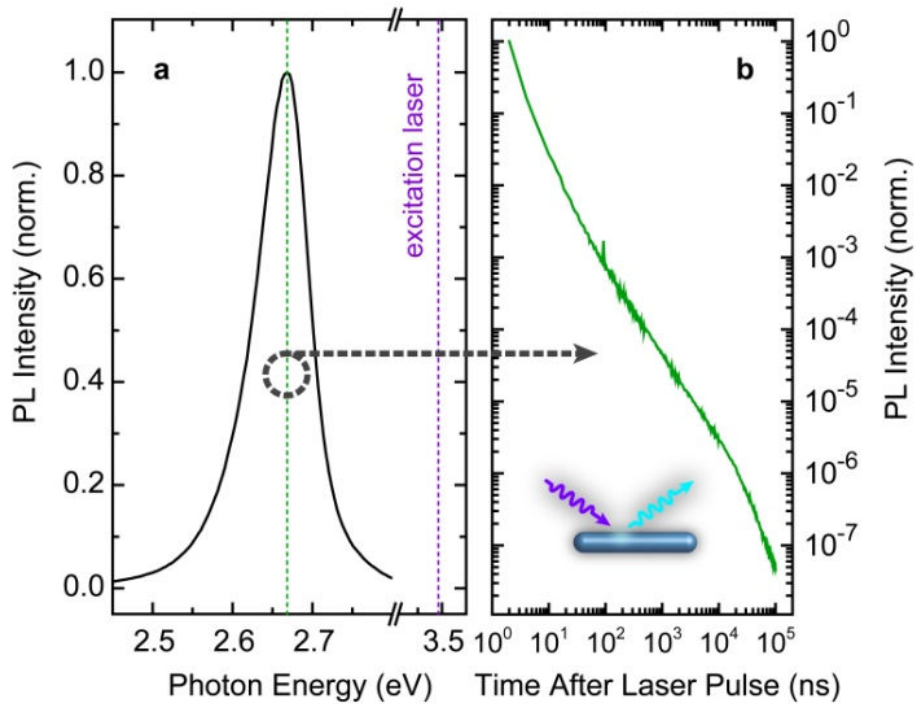

**Figure S1:** a) Prompt band-edge PL spectrum of CdS nanorods at 21 K following excitation with a 355 nm (3.493 eV) laser pulse. b) The emission peak is monitored as a function of delay time from excitation using a gated ICCD camera and spectrometer. The peak emission decay approximately follows a power law, revealing the presence of long-lived "trap" states, which are energetically close to the semiconductor band edge where the exciton forms.

**Correlation of resonances**

To determine which of the three resonances seen in the tetrapods and nanorods (Fig. 3) correspond to a coupled pair of carriers, the time dynamics of the resonances are considered. The correlation of the features in time dynamics in Fig. 3 is independent of temperature and laser power, although both of these parameters directly affect the transient response. The biexponential time dynamics shown in Fig. 3b,d) are characteristic of a (electron-hole) pair process, which has been investigated extensively in the context of conjugated polymers[3].

While the observation of identical dynamics (Fig. 3) alone is sufficient to conclude that each of these paramagnetic centers belong to the same coupled system[3], further proof derives from a comparison of the areas of the two features. Since the area of each resonance represents the probability of inducing a spin transition which causes an optical activity, separately resonant carriers belonging to the same excitation (e.g. an electron and a hole in a pair) must exhibit equal probabilities for this process to occur. To aid in the analysis of comparing the equality of these probabilities, a fitting routine employing three Gaussians was utilized to study the transient spectra. For the CdS nanorod band-edge emission, the results of this analysis are shown in Figure S2. A correlation of the $g \sim 2.00$ and $g \sim 1.84$ resonances is clear since the fitting routine finds comparable areas for the Gaussians representing these two features over a wide range of times. Additionally, we note that the central $g \sim 1.95$ resonance must represent a carrier state which is completely decoupled from the neighboring resonances since it displays marked differences in both probability (i.e. area of the resonance) and time dynamics.

A further point must be made about the disparity between the $T_2$ times given for each of these carriers in Fig. 4) in the main text. The results of the Hahn echo experiment (outlined below) on the $g \sim 1.84$ center of the CdS nanorods give a phase coherence time which is nearly half that of the $g \sim 2.00$ center, as would be expected for a carrier experiencing a larger degree of spin-orbit coupling. The inequality between $T_2$ times of the two (correlated) carriers reflects the unique chemical environments of each and does not conflict with the assignment of the two centers as representing a coupled pair. In fact, and although not measured explicitly, the only hard requirement imposed on the spin states of the pair is that each are characterized by identical $T_1$ times, which is inferred from the equal time dynamics of each resonance[3].

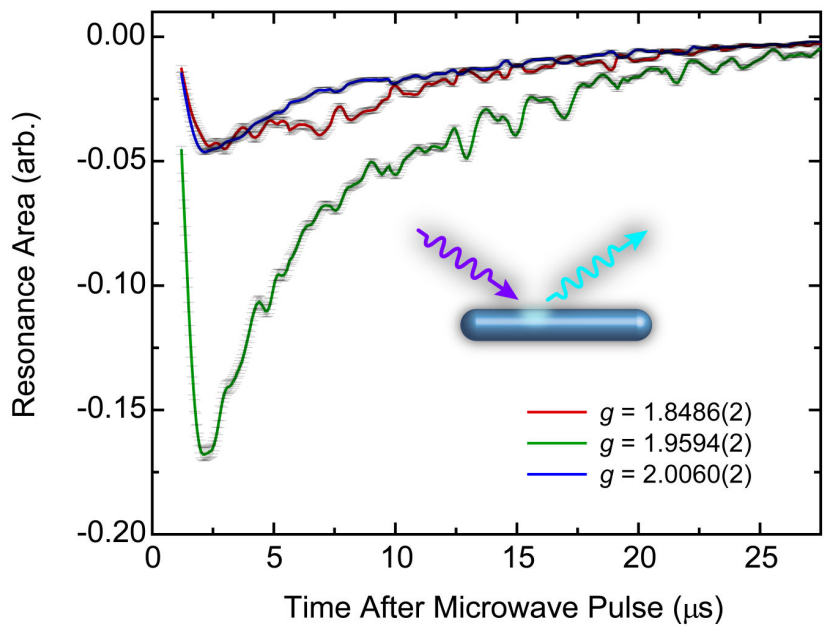

**Figure S2:** Integrated resonances (areas) as a function of time obtained from the triple Gaussian fit applied to the ODMR mapping of CdS nanorod band-edge emission. The equal resonance areas for the $g \sim 2.00$ and $g \sim 1.84$ sites denote the equal probabilities of inducing optical activity following a microwave-induced spin transition. Since the probabilities of inducing such a transition for each of the two sites are equal, it can be concluded that they represent a coupled pair of trap states (i.e. weakly-bound electron-hole pair). The $g \sim 1.95$ state is clearly unrelated.

**Spin identity of trap states and mutual interactions in carrier pairs**

Determining spin identity is a crucial step in chemical fingerprinting as it can help to ultimately illuminate the chemical nature of a trap state for a specific carrier. For example, the complementary knowledge of spin multiplicity, resonance *g*-factor and resonance structure can help to establish the exact symmetry of a paramagnetic site and therefore the exact environment of the localized carrier.

The spin identity of a paramagnetic center can be confirmed in a straightforward manner by carrying out a Rabi nutation experiment since the carrier's spin quantum number is directly reflected in the frequency of oscillation between mutual spin configurations. For transitions between Zeeman-split $m_s$ levels of the form $|S, m_s - 1\rangle \rightarrow |S, m_s\rangle$, and neglecting any significant detuning from resonance, the Rabi frequency is determined by[4] $\Omega_R = \sqrt{S(S+1) - m_s(m_s - 1)} \cdot \gamma B_1$, where $\gamma = \frac{g\mu_B}{\hbar}$ is the gyromagnetic ratio for the center, $\mu_B$ is the Bohr magneton, $\hbar$ is Planck's constant, and $B_1$ is the microwave-induced magnetic field strength at the sample position within the microwave resonator. The *g*-factor is experimentally determined by the resonance center, but once the Rabi nutation has been recorded, a precise value for $B_1$ must be obtained in order to confirm the spin multiplicity of the trap site.

An additional material serving as a standard paramagnetic center with known *g*-factor and spin can be loaded into the microwave resonator alongside the material of interest; in this case, phosphorus-doped crystalline silicon (Si:P with a doping concentration of $[^{31}P] = 10^{16} \, cm^{-3}$). The variations in microwave-induced magnetic field within the resonator volume that contains the combined sample are negligible over the few millimeters of sample breadth, allowing for the direct determination of $B_1$ fields experienced at the trap sites of the nanorods through recording of the Rabi frequency of $^{31}P$ centers in Si.

Aside from establishing the spin identity of the $g \sim 2.00$ and $g \sim 1.84$ sites, we are also interested in the type of mutual interactions experienced by the two carriers. Again, by scrutinizing the frequency components of the Rabi oscillations, general statements can be made as to the prevailing nature of intra-pair coupling. The on-resonance spin-½ system precesses at a frequency of $\Omega_R = \gamma B_1$. As additional, non-negligible interaction terms are introduced into the Hamiltonian describing the spin pair, further frequency components mix with $\Omega_R$ which directly correspond to specific forms of interactions. It has previously been shown[5] that increasing exchange interactions leads to frequency components of $2\gamma B_1$ appearing in the Rabi flopping signal, while an increase in dipolar interactions results in components of[6] $\sqrt{2}\gamma B_1$.

On the other hand, the same frequency components may arise not due to any particular pair interaction, but merely from spin transitions being stimulated within a particular spin manifold. For example, a $\left|\frac{3}{2}, -\frac{1}{2}\right\rangle \rightarrow \left|\frac{3}{2}, \frac{1}{2}\right\rangle$ transition will produce a Rabi frequency component of $2\gamma B_1$, while a strongly exchange-coupled spin-½ system can do the same. This approach of attributing a systematic cause to a measured frequency component is made ambiguous if both the spin identity and the interaction type remain unresolved for the paramagnetic center. There is, however, one case where this ambiguity is easily resolved, which is for the spin-½ pair experiencing weak exchange and dipolar interactions. In this case, the only frequency component present in the Rabi nutation is $\gamma B_1$, which is the case at hand.

Shown in Fig. S3 are Fourier transforms of the Rabi oscillations in Fig. 4a) ($g \sim 1.84$) and Fig. 4b) ($g \sim 2.00$) of the main text. We focus solely on data obtained from the CdS nanorods to investigate the spin state and any possible interactions within the pair since the observed ODMR intensities of the nanorods are an order of magnitude larger than the same transitions seen in the tetrapods. The absence of any additional frequency components in the Fourier spectrum besides the $\gamma B_1$ fundamental implies that this spin-dependent transition results from a pair of weakly-coupled spin-½ carriers, where both exchange and dipolar couplings are negligible. This weak coupling is not beyond expectations for such localized carriers since the distribution of trap sites over the nanoparticles should be random in space, leaving an average pair separation too large for either sufficient wavefunction overlap (exchange) or magnetic dipole-dipole interactions. Such weakly-bound precursor states, which ultimately feed into tightly-bound band-edge excitonic states, are common amongst a variety of material systems, such as hydrogenated amorphous silicon[6] and organic semiconductors[7] and can be manipulated through ESR in order to predetermine the permutation symmetry of final tightly-bound states.

In this case, the carriers comprising this weakly-bound precursor state are each spin-½, which means that they form mutual spin states that can be characterized as either singlet or triplet. This holds for the trapped carriers only, as the band-edge excitonic states are well known to have a higher spin-multiplicity[8].

Therefore, upon detrapping, the singlet/triplet nature of the trapped pair will be projected upon the five individual spin states which make up the exciton fine structure. Since three of these states are bright and two are dark (i.e. spin allowed and forbidden optical transitions), changing the singlet/triplet nature of the trapped carriers will change the probability of moving back into a bright or dark state after detrapping occurs, thereby changing the overall exciton state populations.

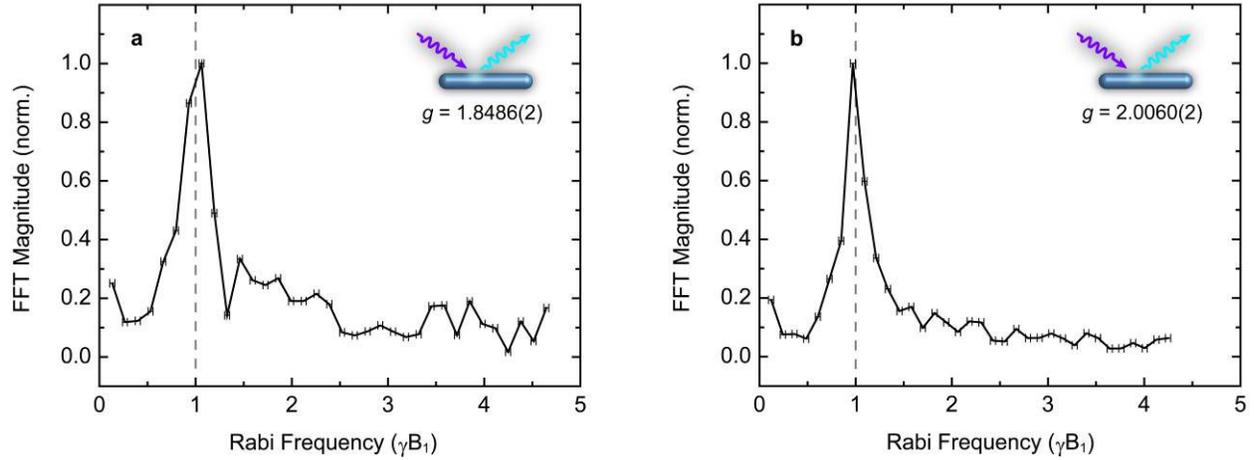

**Figure S3:** Frequency components of Rabi oscillations for both the $g \sim 1.84$ (a) and $g \sim 2.00$ (b) resonances of CdS nanorods. The frequency axis is scaled to $\gamma B_1$, the Rabi frequency of a spin-½ paramagnetic center (marked by the vertical dashed line). There are no additional frequency components, demonstrating that each of these carriers is a spin-½ species and that the coupled pair of carriers experiences negligible exchange or dipolar coupling.

**Measuring spin coherence with Hahn echoes**

A lower limit on the spin dephasing time, $T_2^*$, of a paramagnetic center can be obtained by considering the amplitude decay of the Rabi oscillations. There are two primary mechanisms which artificially shorten coherence time in our system. One is due to the slight inhomogeneities in the oscillating magnetic field of the microwave radiation across the sample, $B_1$, which leads to a distribution of Rabi frequencies, $\Delta\Omega_R$. Another is due to the distribution of local nuclear magnetic moments perturbing the static magnetic field, $B_0$, experienced by the trapped carriers. This distribution leads to an additional detuning term in the Rabi frequency, further increasing $\Delta\Omega_R$. This spread in frequencies evolves the system towards incoherent transitions between the two spin configurations more quickly, but can be overcome by taking advantage of microwave pulse techniques to reveal the true dephasing time of the system, $T_2$. The Hahn echo pulse sequence is particularly appropriate[9]. Since the observable in ODMR is permutation symmetry (i.e. bright or dark mutual spin configuration) and not polarization as in traditional magnetic resonance, we use a slightly modified version of this classic technique. A simple $\frac{\pi}{2}-\tau-\pi-\tau-\frac{\pi}{2}$ pulse sequence is illustrated schematically in Fig. S4a), where a $\pi$-rotation denotes a complete reflection in permutation symmetry for the system and is determined by the precession time measured in a Rabi oscillation. The dynamics involved are straightforward. The first $\frac{\pi}{2}$-pulse places the initially bright spin population into a superposition of bright and dark states. After a delay time, $\tau$, in which the system dephases according to the distribution in Larmor frequencies arising from field inhomogeneities, a $\pi$-pulse is applied in order to reverse the Larmor precession of the system. This reversal effectively takes advantage of the time-reversal symmetry enforced by the long-time stability of the perturbing fields. The subsequent rephasing, or reversal in dephasing, takes place on a timescale equal to that of the initial delay, making the total dephasing time the system is subjected to $2\tau$. The second $\frac{\pi}{2}$-pulse is then applied in order to bring the remaining spin ensemble back to an observable state. By sweeping the $\frac{\pi}{2}$-pulse following the pulse sequence, a small change in the amplitude of transient response (i.e. the differential PL) is measured. This change is referred to as an echo, whose amplitude directly corresponds to the remainder of the initial population. By repeating this sequence and recording the echo as a function of $2\tau$, the loss of spin coherence is observed in the exponential decay of amplitude. Fig. S4b) illustrates this process by displaying some representative echoes using data for the CdS nanorod $g \sim 2.00$ center reproduced from the inset of Fig. 4b) in the main text.

We note that the uncertainty of the quoted spin coherence time of the $g \sim 2.00$ resonance is higher than that of the $g \sim 1.84$ resonance, not due to improved signal-to-noise in the latter, but because of a complicated interference of the electron spin with nuclear magnetic moments. This substructure to the decay amplitude is apparent in Fig. S4b). It arises due to a phenomenon known as electron spin echo envelope modulation (ESEEM). Amplitude modulations of exponential decay of spin phase coherence would be expected to be present in the case of a finer structure splitting of the already Zeeman-split energy levels. These modulations do not prevent extraction of the decoherence time. With higher sensitivity, ESEEM should allow a precise chemical fingerprinting of the trap site in the future by providing information on local nuclear magnetic moments.

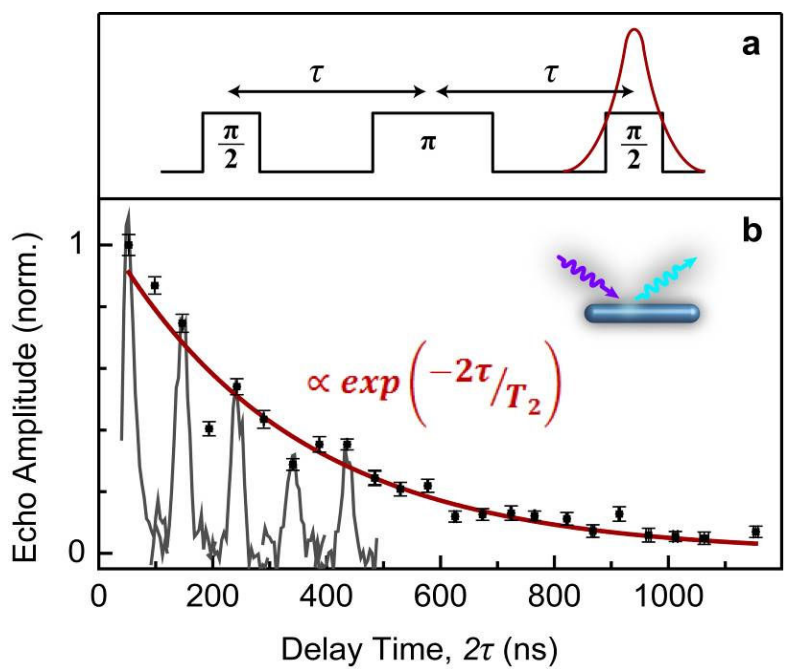

**Figure S4:** Hahn echo pulse sequence to extract the spin dephasing time. A $\frac{\pi}{2}$-pulse projects the dominant initial population into a superposition of bright and dark states. Decoherence due to a distribution in local Larmor precession frequencies is reversed by application of a $\pi$-pulse after a delay time τ. A second $\frac{\pi}{2}$-pulse projects the remaining superposition states (i.e. those which have not lost their spin phase information) back into a bright configuration.

**Dependence of the resonance centers on excitation energy**

In order to probe the energetic distribution of the trap centers, we studied the dependence of the CdSe/CdS tetrapod ODMR spectrum on excitation photon energy. This material system was chosen since excitation could be tuned from above the CdS arm band gap down to the absorption of the CdSe core while monitoring the resonance through the red-shifted PL of the CdSe core. Fig. S5 shows the results of this excitation sequence. The broad, central resonance significantly decreases in amplitude at 488 nm (2.541 eV) excitation compared to 458 nm (2.708 eV) excitation, and disappears completely at 514 nm (2.412 eV). The coupled-pair resonances ($g \sim 2.00$ and $g \sim 1.84$) remain intact, although some broadening is observed with decreasing excitation energy. This observation implies that the species represented by the central resonance has a unique relationship to the delocalized band-edge states of CdS, as compared to the $g \sim 2.00$ and $g \sim 1.84$ coupled-pair species, which exist over a much broader distribution of excitation energies. As commented on in the main text, this observation suggests that the narrow pair species both correspond to CdS surface states which can be populated even by direct excitation of the CdSe core slightly below the CdS band edge. Due to the presence of lattice strain at the heterojunction interface[10], carriers can still become trapped in the CdS even if the excitation energy lies below the band gap of the CdS nanorod. The involvement of lattice strain may explain the slight broadening of the $g \sim 2.00$ and $g \sim 1.84$ coupled-pair resonances with decreasing excitation energy. These resonances are comparatively narrow suggesting that they correspond to discrete atomic sites such as surface defects, organic ligands or the surrounding organic matrix. In contrast, the broad resonance can only be excited when the CdS is pumped above the band edge, suggesting that this species originates from bulk delocalized states in the CdS with substantial disorder broadening due to a wide range of chemical environments probed.

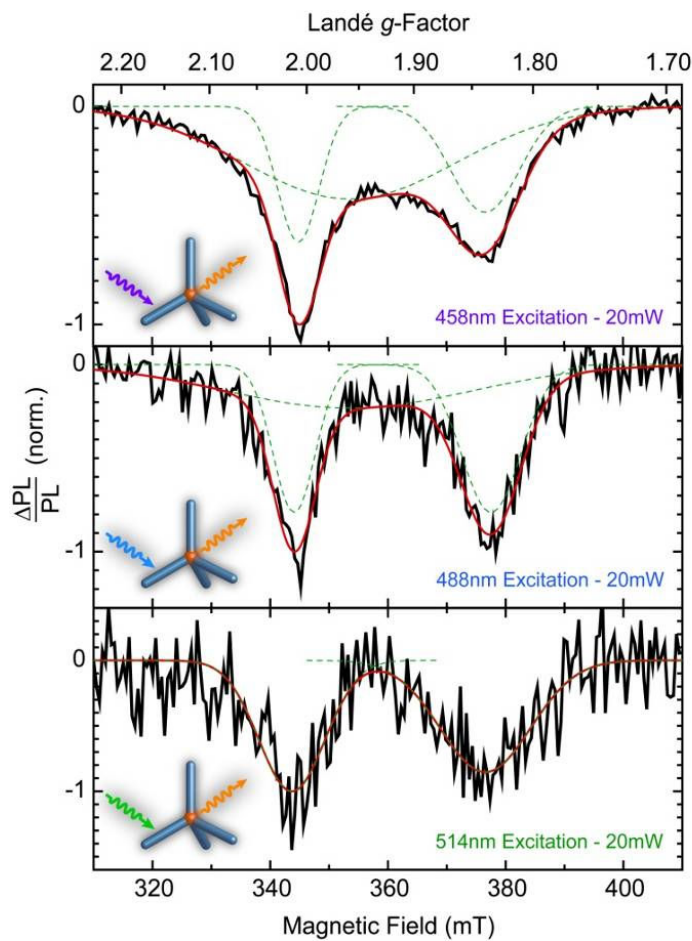

**Figure S5:** Dependence of tetrapod ODMR spectra on excitation energy. As excitation energy is decreased, the central resonance disappears, suggesting a close relationship to band-edge states. The coupled-pair resonances remain, shifting slightly in center position and broadening.

### Discussion on the possible origin of the observed *g*=2 resonance

The resonance present in the CdS nanorods which is most closely aligned with the free-electron *g*-value is that at $g = 2.0060(2)$. Although the same center is observed through the CdSe core emission of the tetrapod structure, since the deep-level chemical defect of the CdS arm also emits at this energy (Fig. 2c), the exact resonance position is likely perturbed due to convolution with the resonance structure of the defect (Fig. 2d). To avoid this convolution we rely on the CdS nanorod data to most accurately assess the features of the $g = 2.0060\,(2)$ resonance, since in this case the excitonic and defect states are spectrally well separated. In describing the nature of the $g \sim 2.00$ resonance, there are two possible models which are supported in the literature. One involves a photogenerated hole becoming trapped at the CdS surface in an undetermined chemical position[11]. Another possibility is that a charge becomes localized to an incorrectly-bonded surface-ligand site[12], or ejected from the nanoparticle into the surrounding organic matrix[13]. Both situations are suspected to constitute a type of charge trap[12]. Each of these situations is expected to result in a resonance position very close to that of the free-electron g-factor ($g \sim 2.0023$).

In the case of the photogenerated hole in CdS, Ref. 10 reported such a site which displayed an axial g-factor asymmetry with $g_{//} = 2.035$ and $g_{\perp} = 2.005$, where parallel and perpendicular refer to the alignment of principal g-factor axes with respect to the external magnetic field, $\vec{B}_0$. For a disordered ensemble of nanocrystals, each of these g-factor axes is randomly oriented with respect to $\vec{B}_0$ and so the spin resonance spectrum will display distinct peaks for each principal g-value, as well as a continuum of peaks between these values representing the linear combination of projections. Such a lineshape is referred to as an anisotropic powder pattern. In general, $g_{\perp}$ results in a higher degree of spin-polarization due to the larger number of axis-normal orientations expressed in the random distribution. This effect would give maximal resonant change in photoluminescence at $g_{\perp}$, as compared to $g_{//}$, which is very near to the situation we observe here.

In considering the second case, that of the charge localized to some organic material (either ligands or matrix), we also find good agreement between our measurements and the expected characteristics for such a material. Due to the extremely low levels of spin-orbit coupling, the g-factor of organic materials is found to be quite close to the free-electron value. Consequently, for a charge which is localized to a surface passivating organic ligand, a resonance very close to $g \sim 2.0023$ would be expected. In addition to the resonance position, the value for the $T_2$ coherence time determined is much longer than those reported for similar inorganic quantum dots[14], yet is of the order of that measured in organic semiconductor systems[14]. Since neither the organic ligands nor the matrix are π-conjugated, it is not presently clear how their chemical structures would support charging, although defect centers respective to these materials are conceivable.

Discrimination between these two models remains difficult at this time without additional information. The level of inhomogeneous broadening and the overlap of the $g \sim 1.95$ resonance presently prevent us from resolving in detail any possible anisotropic features of this resonance. In the previous report of photogenerated holes in CdS[11], detailed information regarding the relative amplitude difference and line width difference between the $g_\perp$ and $g_\parallel$ spectral positions is lacking. We therefore resort to using a single Gaussian line profile in order to represent this resonance as a type of first-order approximation. More parameter information is necessary (line widths and peak intensity ratios) to faithfully make use of a powder pattern fitting function in comparing these two models.

Resolving the ambiguity of chemical assignment for this resonance site could be carried out in at least two ways. One is in using an electron-spin-echo (ESE) detection scheme in order to map out the resonance structure. This technique allows one to independently measure the resonance structure of two overlapping species which have differing coherence times. The ability to separate out the overlapping $g \sim 1.95$ resonance may result in finer resolution of the $g \sim 2.00$ feature details and therefore resolve the issue of line shape anisotropy. A second method of resolving this issue is through taking advantage of the slight amplitude modulation which is likely to be present in the Hahn echo decay of this center, known as ESEEM (described above). By measuring this modulation with higher resolution, both in amplitude and echo delay spacing, the frequency components involved should allow discrimination between the trapped spin interacting with either a local H or Cd nuclear magnetic moment. Such a measurement would give a direct chemical fingerprint of the trap site position.